\documentclass[conference]{IEEEtran}
\ifCLASSINFOpdf
   \usepackage[pdftex]{graphicx}
\else
\fi
\usepackage{ amssymb }

\usepackage[usenames,dvipsnames]{color}

\usepackage{bm}
\usepackage{amsbsy}

\renewcommand{\vec}[1]{\mathbf{#1}}

\begin{document}
%
\title{Using Sparse Gaussian Processes for Predicting Robust Inertial Confinement Fusion Implosion Yields}



%
\author{\IEEEauthorblockN{Peter Hatfield\IEEEauthorrefmark{1},
Steven Rose\IEEEauthorrefmark{2}\IEEEauthorrefmark{1},
Robbie Scott\IEEEauthorrefmark{3}, 
Ibrahim Almosallam\IEEEauthorrefmark{4},
Stephen Roberts\IEEEauthorrefmark{5},
Matt Jarvis \IEEEauthorrefmark{6} \IEEEauthorrefmark{7}
\IEEEauthorblockA{\IEEEauthorrefmark{1} Clarendon Laboratory, University of Oxford, Parks Road, Oxford OX1 3PU, UK}
\IEEEauthorblockA{\IEEEauthorrefmark{2}Blackett Laboratory, Imperial College, London SW7 2AZ, UK}
\IEEEauthorblockA{\IEEEauthorrefmark{3}Central Laser Facility, STFC Rutherford Appleton Laboratory, Harwell Oxford, Didcot OX11 0QX, UK}
\IEEEauthorblockA{\IEEEauthorrefmark{4}King Abdulaziz City for Science and Technology, Riyadh 11442, Saudi Arabia}
\IEEEauthorblockA{\IEEEauthorrefmark{5}Department of Engineering Science, University of Oxford, Parks Road, Oxford, OX1 3PJ, UK}
\IEEEauthorblockA{\IEEEauthorrefmark{6}Astrophysics, University of Oxford, Denys Wilkinson Building, Keble Road, Oxford, OX1 3RH, UK}
\IEEEauthorblockA{\IEEEauthorrefmark{7}Department of Physics, University of the Western Cape, Bellville 7535, South Africa}
}}


\maketitle

\begin{abstract}
Here we present the application of an advanced Sparse Gaussian Process based machine learning algorithm to the challenge of predicting the yields of inertial confinement fusion (ICF) experiments. The algorithm is used to investigate the parameter space of an extremely robust ICF design for the National Ignition Facility, the `Simplest Design'; deuterium- tritium gas in a plastic ablator with a Gaussian, Planckian drive. In particular we show that i) GPz has the potential to decompose uncertainty on predictions into uncertainty from lack of data and shot-to-shot variation, ii) permits the incorporation of science-goal specific \textit{cost-sensitive learning} e.g. focussing on the high-yield parts of parameter space and iii) is very fast and effective in high dimensions.
\end{abstract}


%
\IEEEpeerreviewmaketitle

\section{Introduction}

Inertial confinement fusion (ICF), in which deuterium-tritium (DT) fuel is compressed to temperatures and densities exceeding that found in the Sun, is one of the main potential pathways to nuclear fusion as a source of energy. The world's leading ICF facility is the National Ignition Facility (NIF) at Lawrence Livermore National Laboratory (LLNL). NIF uses \textit{indirect drive} in which lasers first hit a holhraum (typically a gold can), which then emits a thermal radiation field which drives the implosion, in contrast to \textit{direct drive} ICF in which laser beams themselves drive the implosion.

Although huge progress has been made, NIF has been unable to reach the yields originally hoped for. This has led to an interest in using modern machine learning techniques to produce new designs and quantify uncertainties on predictions. \cite{Nora2017} presented an early ensemble of thousands of ICF implosions, and used Gaussian Processes (GPs) to model the parameter space. \cite{Humbird2017} developed a novel neural network (NN) machine learning algorithm called DJINN (`deep jointly-informed neural networks') that used random forests to construct appropriate neural network architectures with relatively little human input, which was used in \cite{Peterson2017} to identify a novel non-spherically symmetric design for NIF. \cite{Humbird2018} used \textit{transfer learning}\footnote{Transfer learning is a family is of machine learning methods that seek to let an algorithm apply information/knowledge gained from one problem, to the task of solving a similar but different problem.} with DJINN to update the machine learning predictions based on seeing real experiments. Finally \cite{Hatfield2019} used genetic algorithms in an even wider parameter space to produce ICF designs almost from scratch.

In this work we use a GP based machine learning algorithm \textit{GPz}, \cite{Almosallam2016a,Almosallam2016b} to build surrogate models of a robust ICF design. GPz has i) a flexible \textit{cost-sensitive learning} feature that permits optimisation for the specific science goal at hand, ii)  models \textit{heteroscedastic noise}, permitting the uncertainty to be decomposed into uncertainty from shot-to-shot variation and uncertainty from lack of data, and iii) uses a sparse framework that lets it run quickly even in high dimensions. We compare it to DJINNs performance on the same data, and discuss future approaches to building ICF surrogates.

\section{Problem Formulation and Methodology}

\subsubsection{The Simplest Design}

The point design for NIF is an indirect drive design, with a capsule of DT gas inside a ablator shell (plastic, CH or a few other options) with a thin layer of solid DT ice on the inside. In the conventional implosion the ablation compresses the DT gas to a relatively high convergence ratio (the ratio of the DT radius at the start of the experiment to the radius at peak compression), and extremely high temperatures. If certain criteria are met, sufficient nuclear reactions take place that alpha heating (heating of the DT plasma by alpha particles produced in the nuclear reactions) dominates, and the gas ignites. This then starts a burn wave propagating through the DT ice layer, from which most of the neutron yield comes from. Most designs use a series of shockwaves to fine-tune the implosion, with several variants studied.

\begin{figure}[!t]
\centering
\includegraphics[width=3.0in]{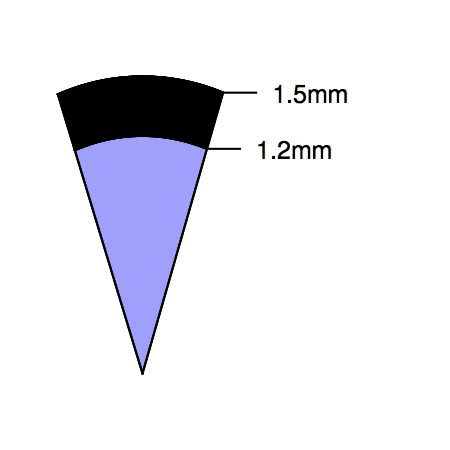}
\caption{Example Simplest Design diagram; black shows CH, pale blue DT gas. Scale and of proportions of the $4 \times 10^{15}$ design of Section \ref{sec_optimal_design} shown, labels are radii from the centre of the capsule.}
\label{fig_sim}
\end{figure}

The Simplest Design removes some of the more complex/challenging aspects of ICF where the physics is more uncertain: i) delicate pulse timing, ii) the burn wave through the DT ice and iii) high convergence ratios. This design simply has DT gas (but of a much higher density), with a CH ablator, and a Gaussian drive (sketch of capsule design shown in figure \ref{fig_sim}). It is unlikely that the Simplest Design will be able to lead to ignition at NIF, but it does represent a pathway to very predictable robust 1D implosions. We use a thermal x-ray drive with a Gaussian temperature time dependence.

We parametrise the Simplest Design with 5 parameters and investigate the parameter space within the following limits:

\begin{itemize}
  \item $50\mathrm{eV}<T_{\mathrm{peak}}<400\mathrm{eV}$ - the peak temperature of the drive
  \item $0.01\mathrm{ns}<\sigma<5\mathrm{ns}$ - the standard deviation of the time dependence of the temperature of the drive
  \item $0.1\mathrm{mm}<r_{1}<1.5\mathrm{mm}$ - the radius of the DT gas
  \item $0.05\mathrm{mm}<r_{2}<1\mathrm{mm}$ - the radius of the CH ablator
  \item $10\mathrm{mg/cc}<\rho<200\mathrm{mg/cc}$ - the density of the DT gas (mg/cc)
\end{itemize}

N.B. the gas fill is much higher than typical for ICF (normally closer to $\sim0.1\mathrm{mg/cc}$) - meaning that the implosions have a much lower convergence ratio, but that we would not expect the design to be able to  reach yields comparable to that which are in principle possible with the point design on NIF. This is because the point design has low density gas that is (comparatively) easily compressed. At peak compression, the hotspot (the low density gas that has been compressed) and DT ice are approximately isobaric, but the density is highly non-uniform - resulting in a small region of high temperature in the hotspot, with low temperature in the ice. The temperatures in the low density hotspot are in principle sufficient to initiate significant burn within the hotspot (which hopefully drives a burn wave through the DT ice, to give extremely high yields). For an equivalent DT mass arranged uniformly (as a function of radius), the energy required to compress it sufficiently to heat it to temperatures high enough to initiate fusion would be far higher than that available on NIF.  The design space is chosen to roughly correspond to what is achievable with a gold hohlraum on NIF; the main restrictions are total capsule radius and total energy in drive.

The Simplest Design shares some design philosophy with the 2-shock design of \cite{Khan2016} and \cite{MacLaren2018}. The design of shot N161004 described in \cite{MacLaren2018} still has a DT ice layer like more typical designs, but has a relatively high gas fill density of 5mg/cc (much closer in log-space to the densities we consider than to the point design), and a correspondingly more modest convergence ratio. They also have a relatively simple 2-shock drive. These features are experimentally observed to make the implosion much more 1D, and much closer to the predictions of simulations. The Simplest Design goes a step further, with an even higher gas fill, even lower convergence ratio and and an even simpler 1-shock drive - and thus also has the potential to give a robust 1D implosion that closely matches simulation. 

\subsubsection{Data}

Typically machine learning methods give better predictions in parts of parameter space with lots of data, and vice-versa. We use here a Monte Carlo sampling, but rather than sampling uniformly in the parameter under consideration, points are sampled from a multi-variate Gaussian\footnote{In log-space} centred on where a preliminary estimate for an optimal design was\footnote{Optimal sampling in the parameter space likely depends on the end goal e.g. if planning to eventually implement transfer learning it may still be valuable to sample the low-yield parts of parameter space.}. \cite{Peterson2017} conversely use latin hypercube sampling (LHS) to achieve good coverage of the parameter space, which may be a valuable alternative sampling approach for future work.  This base design was $T_{\mathrm{peak}}=300\mathrm{eV}$, $\sigma=2\mathrm{ns}$, $r_{1}=1\mathrm{mm}$, $r_{2}=0.15\mathrm{mm}$ and $\rho=55\mathrm{mg/cc}$. 5000 simulations were run and we divided it into 30\% training, 30\% validation and 40\% testing data. The training data is used to infer the large number of parameters that make up the machine learning predictive model, the validation data is used to infer model hyper-parameters (essentially the complexity of the model), and the test data is held back to use as a final test of performance. We simulate shot-to-shot variation by artificially adding Gaussian scatter to the calculated log-yields\footnote{Log-yields always measured in logarithms of base ten of number of neutrons} in number of neutrons. This is chosen to be $\rho$ dependent; the scatter is chosen to have standard deviation $0.1 \times \rho/(55\mathrm{mg/cc})$ dex.

Our implosion simulations are performed using the {\sc Hyades}\cite{Larsen1994} radiation-hydrodynamics simulation code, which is well benchmarked and used widely for the simulation of inertial fusion and high energy density physics applications\cite{Keilty2000,Leibrandt2005,Sequoia2006, Scott2012}. {\sc Hyades} models hydrodynamics within a Lagrangian framework. Electron and ion thermal energy transport is described by a flux-limited Spitzer-H{\"a}rm thermal conductivity model. Equations-of-state either use the Los Alamos SESAME tables\cite{Hu2011} or QEOS\cite{More1988}. Ionization levels come from a hydrogenic average-atom model or self-consistently from QEOS. Radiation transport uses the multi-group diffusion approximation; here we use 60 groups. A 1D spherically symmetric geometry is employed.

Neither laser-plasma interactions nor hohlraum physics are modelled; instead we use an incoming x-ray drive imposed at the outside of the grid. The capsules are modelled within a 5mm helium container. Simulations start in cryogenic conditions at $1.551\times10^{-3}$eV$=18$K. The {\sc Hyades} runs were performed on SCARF at the Central Laser Facility at Rutherford Appleton Laboratory using 500 CPUs. The modelling used here is likely appropriate for the design investigated here\footnote{It is in some sense hard to definitively know this is the case until a real shot has been performed, so this statement must remain conditional without experimental data.  However the aforementioned design of \cite{MacLaren2018} finds that 1D models are in reasonable agreement with the data; and the implosions described here should be `even more 1D'.}, but the next level of sophistication of modelling would be to a) move to 2D/3D simulations (e.g. so Rayleigh-Taylor instabilities are incorporated) and b) start to include laser/hohlraum physics and non-Planckian drives (e.g. simulate the laser light being converted to x-rays, rather than just assuming an incoming x-ray drive). See \cite{Rosen1999,Hamza2005} for overviews of the wide range of physics involved in ICF, and associated issues concerning which physics to include in simulations etc.

\subsubsection{GPz} \label{sec:GPz}

GPz is a machine learning regression algorithm originally developed for the problem in astrophysics of calculating the photometric redshifts of galaxies; the details of the algorithm and the key developments in machine learning (ML) theory are described in \cite{Almosallam2016a,Almosallam2016b}, and applied to photometric redshift calculation in \cite{Gomes2017,Duncan2017}, and to orbital dynamics in \cite{Peng2019}. The algorithm is `GP' based; a Gaussian Process is a stochastic process with a random variable defined at each point in a space of interest, such that any finite subset of the random variables has a multivariate normal distribution (equivalently any linear combination of random variables from different points has a normal distribution). A GP is essentially an un-parametrised continuous function defined everywhere with Gaussian uncertainties. A GP based ML algorithm will typically take a set of data over the parameter space of interest and in some sense try and find the function in the function space defined by the Gaussian process that was most likely to have produced the data - and then make predictions for other parts of parameter space based on that.

GPz is a sparse Gaussian process based code, a fast and a scalable approximation of a full Gaussian Process \cite{Rasmussen2006}, with the added feature of being able to produce input-dependent variance estimations (heteroscedastic noise). For the full details of the algorithm see \cite{Almosallam2016a,Almosallam2016b,Almosallam2017}, but we summarise the main details here. The model assumes that the probability of the observing a target variable $y$ given the vector input $\vec{x}$ is $p(y|\vec{x}) = \mathcal{N}(\mu(\vec{x}),\sigma(\vec{x})^2)$. The mean function, $\mu(\vec{x})$, and the variance function $\sigma(\vec{x})$ are both linear combinations\footnote{N.B. This is a different sense of linear combination to that which can be used in the definition of a Gaussian Process. That definition of a Gaussian Process involves the addition of random variables at different points; the linear combination discussed here is an addition of multivariable functions to build up a representation of $\mu(\vec{x})$ and $\sigma(\vec{x})$ .} of `basis functions' that
take the following form:

\begin{equation} \label{eq:mu}
\mu(\vec{x}) = \Sigma^{m}_{i=1} \phi_{i}(\vec{x})w_i
\end{equation}
\begin{equation} \label{eq:sigma}
\sigma(\vec{x}) =\exp \left( \Sigma^{m}_{i=1} \phi_{i}(\vec{x})v_i \right)
\end{equation}

where $\{\phi_{i}(\vec{x}),w_i,v_i\}^{m}_{i=1} $ are sets of $m$ basis functions and their associated weights respectively\footnote{In practice GPz actually uses $\sigma^{2} = \exp(\Sigma_{i=1}^{m}\phi(\vec{x})v_{i}+b)$ where addition of the bias term `$b$' is used for practical reasons but is not required theoretically. }. Basis function models (BFM), for specific classes of basis functions such as the squared exponential, have the advantage of being universal approximators, i.e. there exist a function of that form that can approximate any function, with mild assumptions, to any desired degree of accuracy; i.e. a one size fits all function. BFM are a form of sparse Gaussian Processes \cite{Almosallam2017}.The most general form of the squared exponential is:

\begin{equation}\label{eq:basis_function}
\phi_{i}(\vec{x})=\exp \left(-\frac{1}{2}(\vec{x}-\boldsymbol{\mu_i})^{T} \Lambda_{i}^{-1 }(\vec{x}-\boldsymbol{\mu_i}) \right)
\end{equation}

The goal of GPz essentially is to find the optimal parameters, $\{\boldsymbol{\mu_i}, \Lambda_{i}, w_i, v_i \}^{m}_{i=1}$ such that the mean and variance functions, $\mu(\vec{x})$ and $\sigma(\vec{x})$, are the most likely functions to have generated the data - using Bayesian inference. These optimal parameters are found with a Limited-memory Broyden-Fletcher-Goldfarb-Shanno algorithm \cite{Nocedal1980} (a hill-climbing optimisation algorithm for differentiable non-linear problems). During the training stage of the algorithm GPz is inferring the locations and spreads that describe the basis functions; during the validation stage it is inferring the appropriate complexity of the model, essentially how many basis functions to use.

The key features introduced by GPz include a) implementation of a sparse GP framework, allowing the algorithm to run in $O(nm^2)$ instead of $O(n^3)$ (where $n$ is the number of samples in the data and $m$ is the number of basis functions), b) a `cost sensitive learning' framework where the algorithm can be tailored for the precise science goal, and c) properly accounts for uncertainty contributions from both variance in the data as well as uncertainty from lack of data in a given part of parameter space (by marginalising over the functions supported by the GP that could have produced the data).

\begin{figure}[!t]
\centering
\includegraphics[width=3.0in]{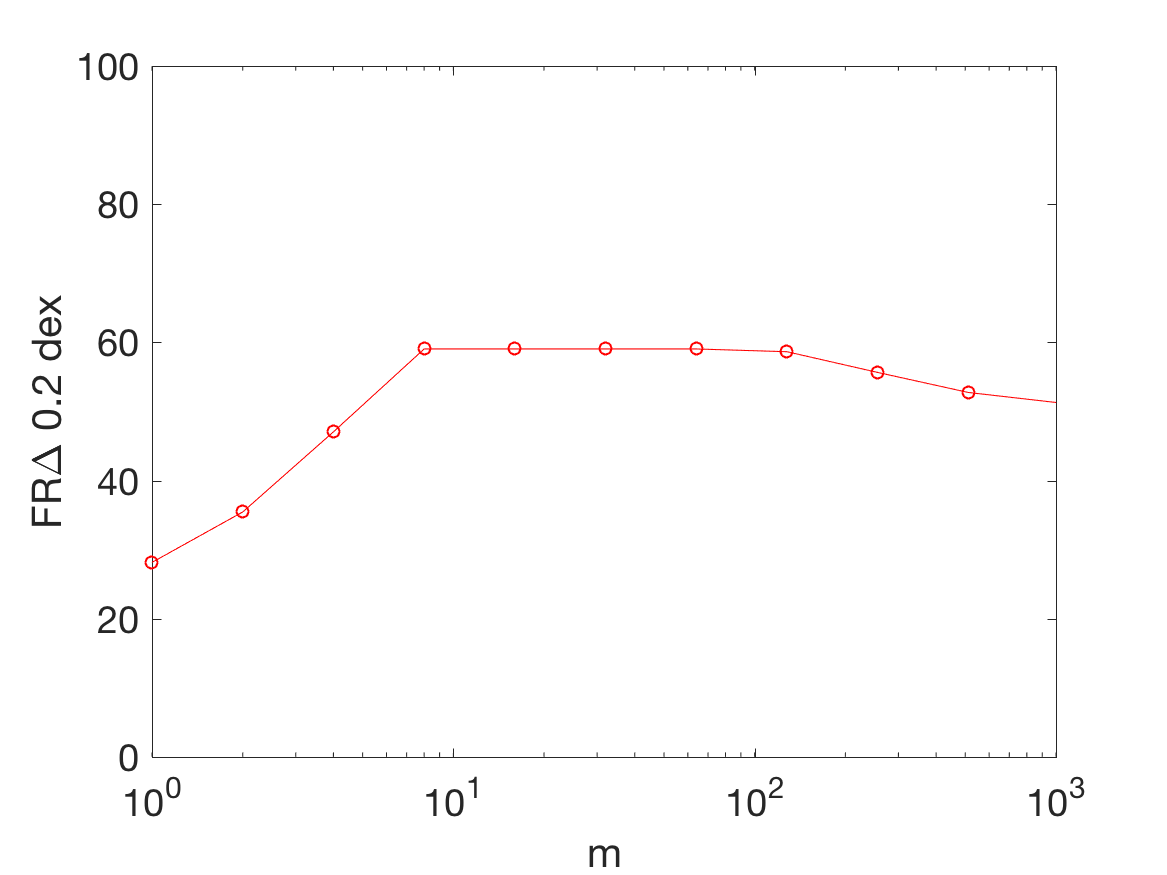}
\caption{The performance of the algorithm as a function of the number of basis functions used. The y-axis is the fraction of test data within 0.2 dex of the prediction.}
\label{number_basis}
\end{figure}

Unless otherwise stated, we use the settings in Table \ref{table-settings} (see \cite{Almosallam2016a,Almosallam2016b} for precise definitions and interpretations). GPz requires very little fine-tuning. The most important parameter is $m$, the number of basis functions. A higher $m$ corresponds to higher model complexity and longer training times. Figure \ref{number_basis} shows algorithm performance as a function of $m$; best performance is achieved for $m\approx 10-100$, in line with findings in \cite{Almosallam2016a,Almosallam2016b}.

 \begin{table*}
\caption{Parameter setting of GPz.}
\begin{center}
\begin{tabular}{| l | l | l |}
    Parameter 	&	Value		&	Description\\	\hline
	$m$			&	100		&	Number of basis functions (the $\phi_i$); complexity of GP\\
	maxIter		&	500		&	Maximum number of iterations (comparisons with the validation data) permitted\\
	maxAttempts	&	50		&	Maximum number of iterations to attempt if there is no progress on the validation set\\
	method		&	GPVC	&	Type of bespoke covariances (the $\Lambda_i$ in equation \ref{eq:basis_function}) used on each basis function (see \cite{Almosallam2016b} for the different options)\\
	normalize	&	True	&	Pre-process the input by subtracting the means and dividing by the standard deviations\\
	joint		&	True	& Jointly learn a prior linear mean-function (learn the function means and variances jointly)\\
	heteroscedastic	 & True & Model noise as well as point estimates
  \end{tabular}
\end{center}
\label{table-settings}
\end{table*}

GPz permits custom cost-sensitive learning e.g. a specific science goal to specified to the algorithm. In the application of calculating galaxy redshifts this typically might be something like specifying that low-redshift galaxies are a low priority. In the ICF case, we can also set particular parts of parameter space of interest e.g. areas around a cliff, parts of parameter space that are possible with a particular facility etc. In this work we will use the weighting $w=\sqrt{Y}$ (where $Y$ is the simulated yield in number of neutrons), up-weighting the cost of getting predictions wrong in the high-yield part of parameter space, and assuming that we are relatively insensitive to getting predictions wrong in low-yield parts of parameter space. The choice of  $w=\sqrt{Y}$ was motivated by requiring a function that was a) monotonically increasing with $Y$ (so that higher yield parameter space has a higher weight than lower yield parameter space), b) a power law (so that the weighting is invariant under rescaling of $Y$) and c) has a `reasonable' dynamical range for the design space considered (here yield spans $\sim10^{10}-10^{16}$, so the ratio of weightings of different parts of parameter space can reach up to $\sim \sqrt{\frac{10^{16}}{10^{10}}}=1000$).  

The 5D space considered here is of relatively modest dimensionality, but GPz has been shown to be effective and fast running in $\gtrsim$10D, see \cite{Gomes2017} and subsection \ref{sec_scale}.

We also trialled our data with DJINN \cite{Humbird2017}\footnote{Taken from the version on \textit{https://github.com/LLNL/DJINN}}. The DJINN solution to difficulties in designing neural network structure is to use a novel mapping from decision tree to network structure, giving a very user-friendly algorithm that works with very little human input in a wide variety of circumstances. Parameters used for DJINN used are shown in table \ref{table-settings-DJINN}, with choices motivated by \cite{Humbird2017} (improved performance may be achievable with further parameter fine-tuning).

 \begin{table*}
\caption{Parameter setting of DJINN.}
\begin{center}
\begin{tabular}{| l | l | l |}
    Parameter 	&	Value		&	Description\\	\hline
	ntrees			&	10		&	Number of neural nets in ensemble\\
	maxdepth		&	5		&	Maximum depth of tree\\
	dropoutkeep	&	0.95		&	Dropout \\
	niters		&	100		&	Number of times network evaluated for\\
  \end{tabular}
\end{center}
\label{table-settings-DJINN}
\end{table*}

\section{Results}

\subsubsection{Predictions and Optimal Design} \label{sec_optimal_design}

Figure \ref{fig_comparison} shows the yield predicted by the machine learning algorithms for the test data, compared with the Hyades yields. GPz is largely able to correctly predict the yield for most of the test data, with reasonably realistic uncertainties. Figure \ref{fig_param_space} shows the yield as a function of two of the parameters and the corresponding uncertainty on the predictions. The part of parameter space with maximal yield within design constraints (e.g. what is feasible for a given facility) can easily be extracted, or a similar stability test as in \cite{Peterson2017} can be used to find the design with the best combination of yield and stability etc. For example, say we fix the drive at $T_{\mathrm{peak}} = 300\mathrm{eV}$ and $\sigma = 1.5$ns, and restrict interest to capsules smaller than 1.5mm (e.g. $r_{1}+r_{2}<1.5$mm). The capsule with highest predicted yield is easily found to be $r_{1}\approx1.2$mm, $r_{2}\approx0.3$mm and $\rho\approx25$mg/cc, giving a yield of $Y \sim 4 \times 10^{15}$. However suppose we were only interested in making sure that the design robustly had a yield above $Y =10^{15}$, and instead wanted to minimise the capsule radius. We define $P(Y>10^{15}|\delta=0.2)$  as the fraction of designs that still have a yield above $Y =10^{15}$ when each parameter is perturbed by an amount sampled from a Gaussian with a standard deviation of 20\% (c.f. \cite{Peterson2017}). We can find the design that minimises $r_{1}+r_{2}$ with $P(Y>10^{15}|\delta=0.2)>0.9$, leading to a slightly different design, $r_{1}\approx0.95$mm, $r_{2}\approx0.3$mm and $\rho\approx25$mg/cc  ($Y \sim 2.5 \times 10^{15}$).

\subsubsection{Cost-Sensitive Learning}

Figure  \ref{fig_comparison} shows a comparison of GPz  $w=1$ (`normal' in \cite{Almosallam2016a}) and using cost-sensitive learning ($w=\sqrt{Y}$). Figure \ref{fig_bias} and \ref{fig_RMSE} show the bias (mean of $Y_{\mathrm{test}}-Y_{\mathrm{prediction}}$) and root-mean-squared-error (RMSE) on the predictions, illustrating that performance in higher yield parts of parameter space (log$Y\sim15-17$) is indeed improved at the cost of performance in lower yield regions. In general CSL can be linked to specific science goal of a study e.g. for the goal of achieving ignition at the NIF we probably have particular interest in having low RMSE close to cliff edges, but are relatively insensitive to RMSE or bias both far above and far below this boundary. We would note that CSL is a method that can in some circumstances obtain slightly better statistical properties in certain parts of parameter space \cite{Duncan2017}; but it cannot extract information from the data that simply isn't there (e.g. a really extreme weighting scheme will still fail to improve predictions in parts of parameter space with almost no data).

\subsubsection{Comparison with DJINN}

\begin{figure*}[!t]
\centering
\includegraphics[width=8.0in]{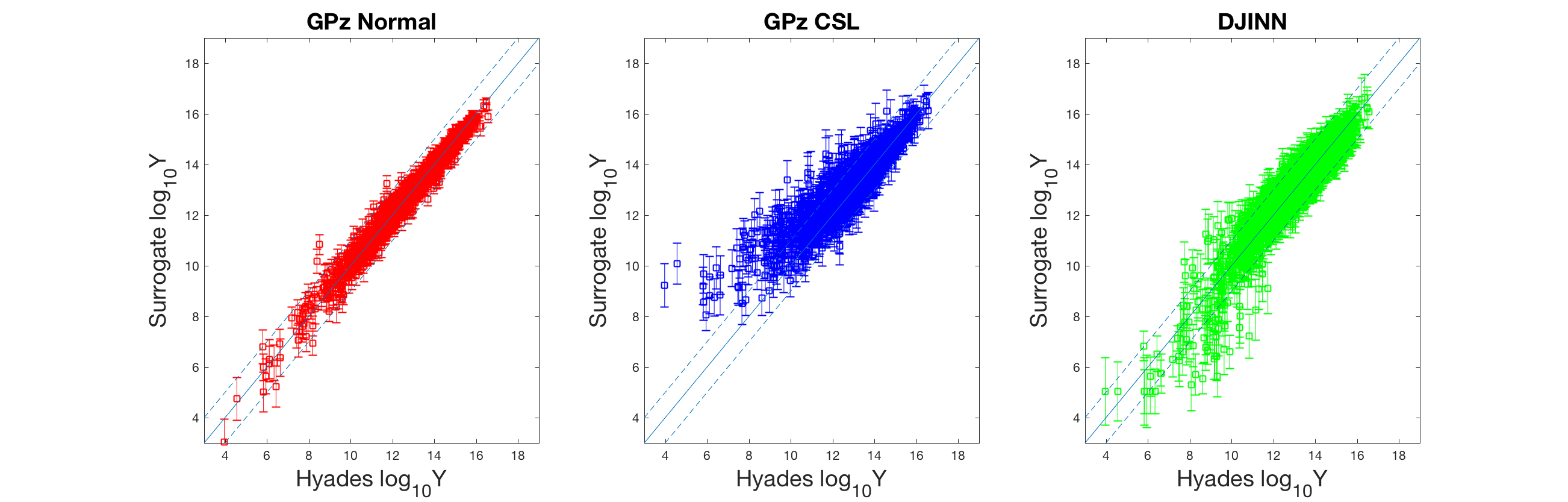}
\caption{ Surrogate yield and Hyades yield compared for different surrogate set-ups.. GPz-Normal is shown in red (left), GPz-CSL in blue (centre), DJINN in green (right). The diagonal filled line shows equality, and the dashed lines show one dex discrepancies}
\label{fig_comparison}
\end{figure*}

\begin{figure}[!t]
\centering
\includegraphics[width=3.0in]{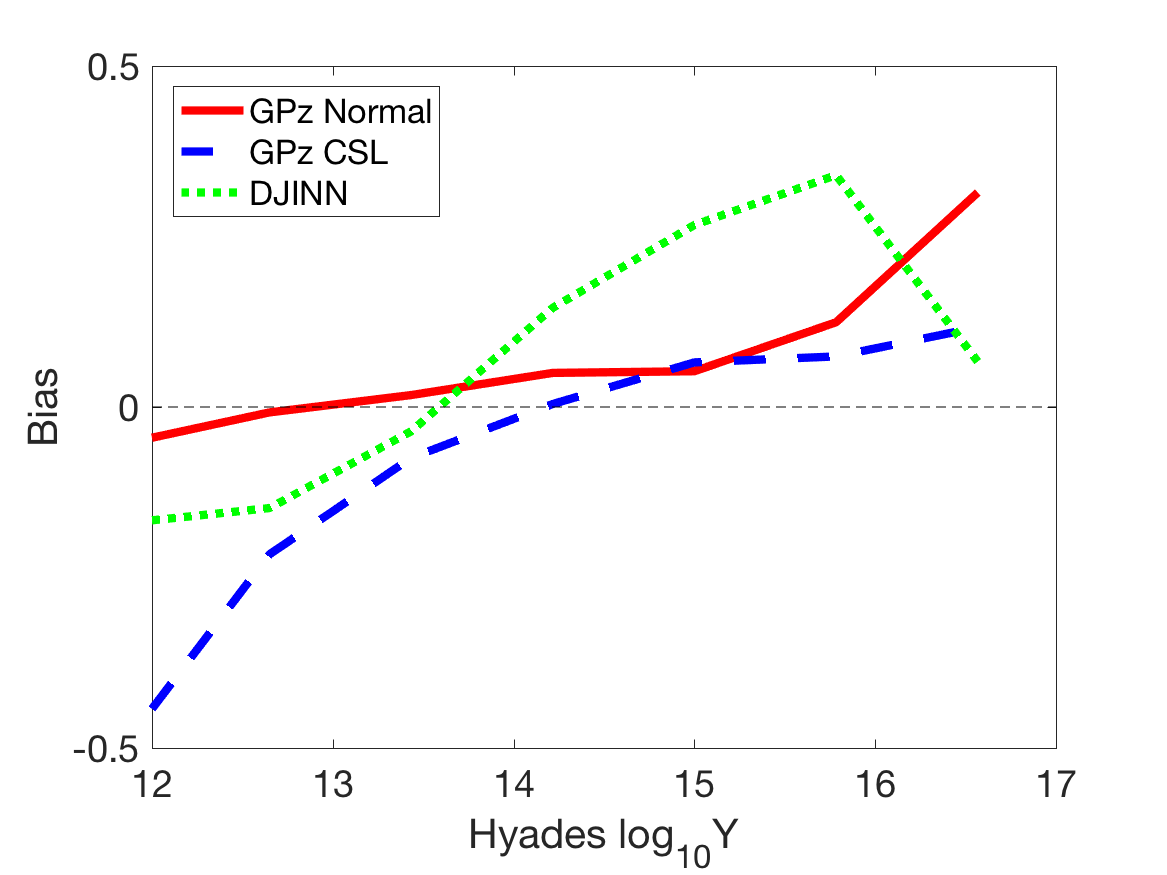}
\caption{The bias on the predictions, GPz-Normal shown in red (solid line), GPz-CSL in blue (dashed), DJINN in green (dotted).}
\label{fig_bias}
\end{figure}

\begin{figure}[!t]
\centering
\includegraphics[width=3.0in]{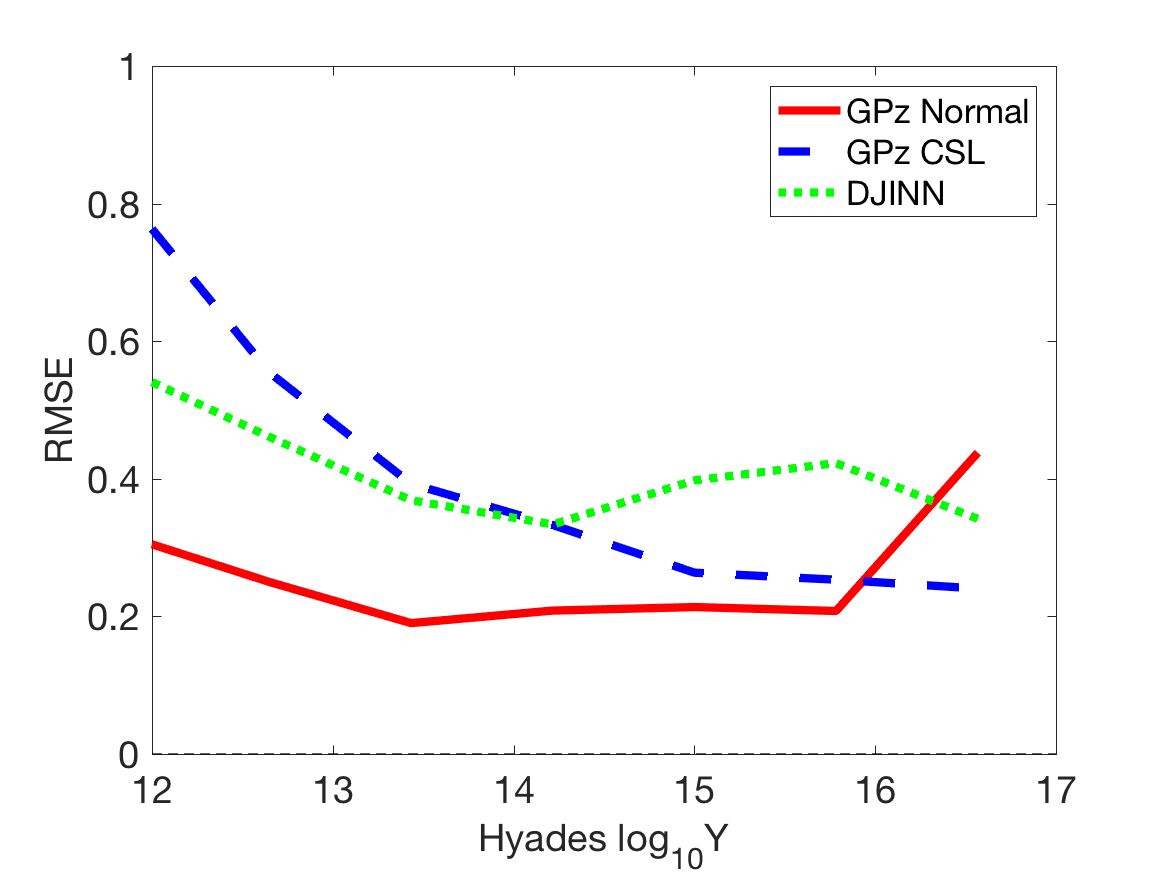}
\caption{The RMSE on the predictions, GPz-Normal shown in red (solid line), GPz-CSL in blue (dashed), DJINN in green (dotted).}
\label{fig_RMSE}
\end{figure}

We also applied (Bayesian) DJINN to our data, also shown in Figures \ref{fig_comparison}, \ref{fig_bias} and \ref{fig_RMSE}. The intention is not to do a rigorous code comparison, as the codes were developed for different machine learning goals and it is non-trivial to directly compare model complexity, but simply to illustrate that Gaussian processes are comparably viable tools for building ICF surrogates.

DJINN also performs well in producing a surrogate model. The algorithm trained and predicted in $\sim$80s and $\sim$2s respectively, compared with $\sim$5s and $\sim$0.02s for GPz on a laptop\footnote{16 GB RAM, 3.1 GHz Intel Core i5}.

Some authors have found that the best results are achieved using a committee of a variety of machine learning methods, so it is possible best results could be achieved using a combination of neural nets and Gaussian processes with an ensemble averaging \cite{Clemen1989} or hierarchical Bayesian approach \cite{Duncan2017}. This would also start a move towards non-Gaussian pdfs, as predictions near cliffs are likely to be multi-modal (an experiment either ignites or it doesn't). Greater precision in key parts of parameter space can of course also be achieved by doing more simulations in that part of parameter space, but that requires advance knowledge of which part of parameter space is interesting. Future work could couple the process of sampling parameter space and building the surrogate. It might also be interesting to consider how CSL-like methods might be implemented within DJINN e.g. loss-calibrated learning as per \cite{Cobb2018}.

\subsubsection{Uncertainty Decomposition}

\begin{figure*}[!t]
\centering
\includegraphics[width=7.5in]{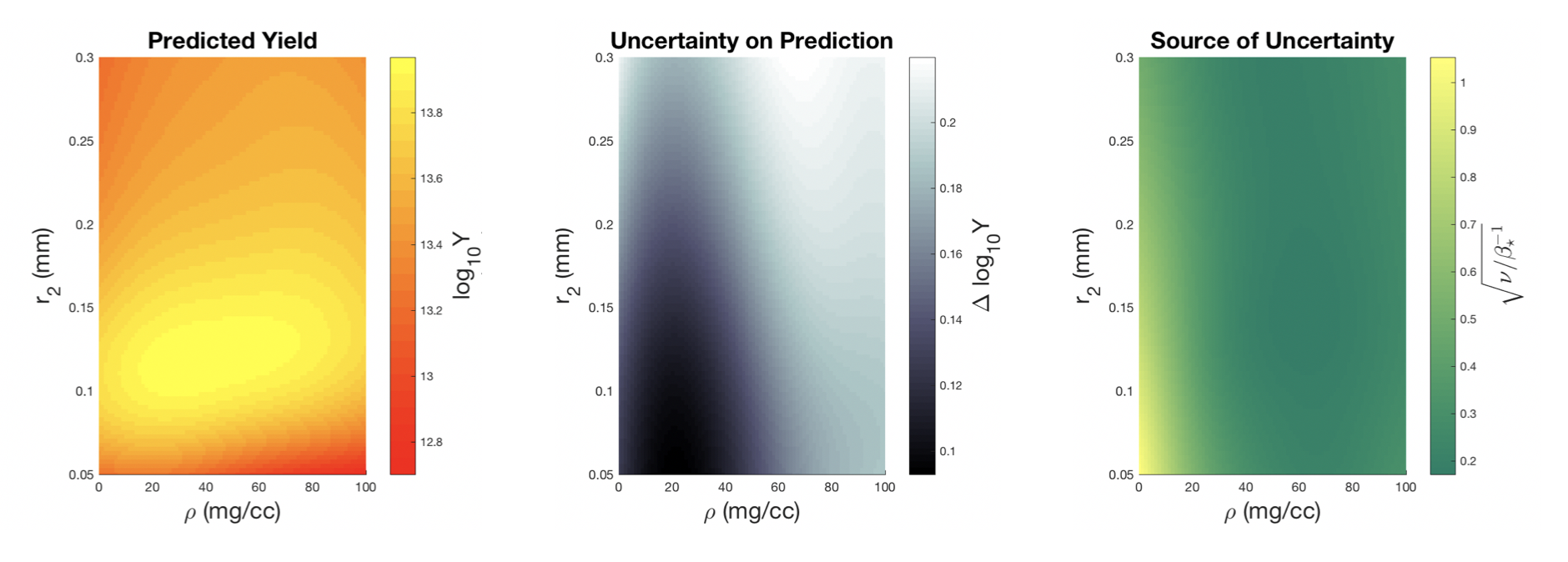}
\caption{A 2D slice through the 5D parameter space (with the other three parameters set to $r_1=1$mm, $T_{\mathrm{peak}}=300$eV, $\sigma=0.5$ns), showing the surrogates yield (left), the uncertainty on the prediction (centre) and the uncertainty from lack of data divided by uncertainty from shot-to-shot variation (right).}
\label{fig_param_space}
\end{figure*}

The right subplot of Figure \ref{fig_param_space} shows $\sqrt{\nu/\beta_{\star}^{-1}}$, the ratio of uncertainty from lack of data to uncertainty from intrinsic variation. $\nu$ is the variance from lack of data, defined as $\nu=\boldsymbol{\phi}(\vec{x})\Sigma^{-1}\boldsymbol{\phi}(\vec{x})^{T}$, where $\boldsymbol{\phi}$ is the vector of non-linear basis functions that the prediction mapping is constructed from, $\vec{x}$ is the test data, and $\Sigma$ is the covariance matrix of uncertainty on the weights applied to the basis functions when constructing the posterior mean. It is essentially the uncertainty on the $\mu(x)$ in equation \ref{eq:mu}. $\beta_{\star}^{-1}$ is the input-dependent noise variance. It is essentially the $\sigma(\vec{x})^{2}$ in equation \ref{eq:sigma}.  The total variance is $\sigma_{\mathrm{Total}}^2=\nu + \beta_{\star}^{-1}$.  See Equations 3.13 and 5.10 in \cite{Almosallam2016b} for a more in depth explanation of the calculation and interpretation of these quantities, and \cite{Bishop2006} for a more general background. The plot illustrates that GPz correctly identifies that for $\rho<10$mg/cc more of the uncertainty is coming from lack of data rather than shot-to-shot variation (as there were no simulations done in that part of parameter space). This shows that GPz can help understand what is the dominant source of uncertainty in different parts of parameter space. This uncertainty decomposition does however come with the caveat that uncertainty is still likely underestimated at the edge of the domain, and anywhere far from data, due to the use of only a finite number of basis functions. Estimates of shot-to-shot variation can also be incorporated into design optimisation e.g. suppose we wanted to find the design with a yield above $Y =10^{15}$, but with minimal shot to shot variation, we can find the design with smallest $\beta_{\star}^{-1}$ that meets this yield criterion, which is $r_{1}\approx1.05$mm, $r_{2}\approx0.15$mm and $\rho\approx15$mg/cc  (for the same drive as in Section \ref{sec_optimal_design}.

The shot-to-shot variation considered here is not quite identical to the real problem in ICF; here we added scatter to the thousands of simulations, whereas more realistically we might have a large number of simulations with no scatter/uncertainty, and then just a few experiments with some shot-to-shot variation. Nonetheless an approach similar to that detailed here could be effective in folding shot-to-shot variation into surrogate building. For example, one might simulate a large number of shots varying both parameters the designer controls (e.g. capsule shell thickness) as well as ones they don't (e.g. imperfections on the surface of the shell). The outputs from these simulations could then be used to build a `noisy' surrogate as a function of the controlled parameters. Another useful feature is that understanding the shot-to-shot variation permits the model to say that there is substantial amount of uncertainty in the prediction of the yield for an individual shot, but also to say that there is no need to do any further experiments in that part of parameter space as all the uncertainty is from shot-to-shot variation. Features of GPz in development that could be useful for ICF surrogate building in future include i) incorporating uncertainty on predictions due to uncertainty on input parameters (e.g. if there are error bars associated with the gas fill densities and shell thicknesses etc.) and ii) coping with incomplete data (e.g. training on data where a subset of the shots don't have some of the experimental properties recorded).

\subsubsection{Scalability} \label{sec_scale}

\begin{figure}[!t]
\centering
\includegraphics[width=3.0in]{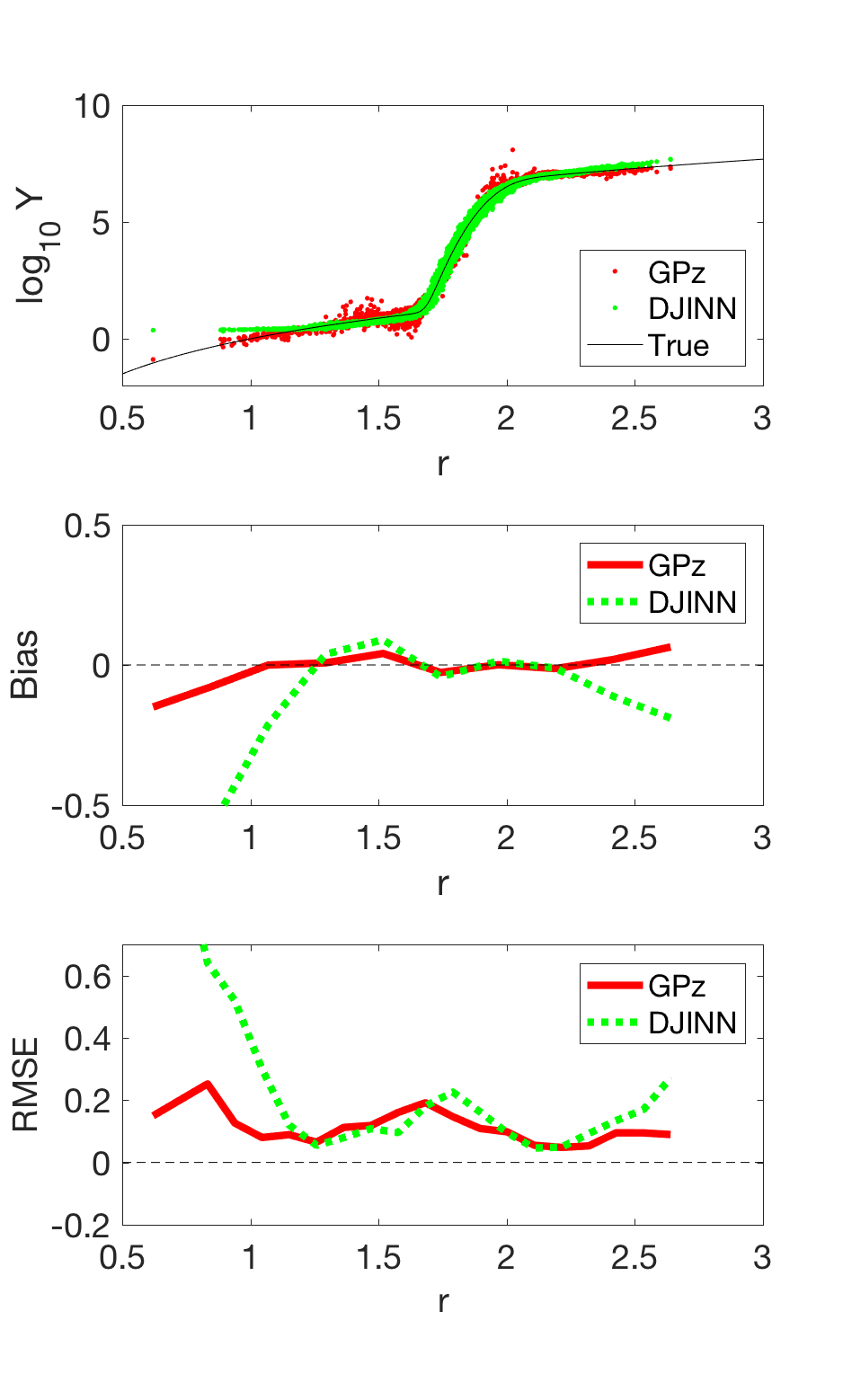}
\caption{Top plot: GPz (red) and DJINN (green) predictions as a function of radius for our analytic model, with the analytic result shown with a black line. NB. these are results in 10D; we plot the 1D $r$ and $Y$ values of the test data for ease of visualisation. Middle plot: RMSE on the predictions, GPz-Normal shown in red (solid line), DJINN in green (dotted). Bottom plot: RMSE on the predictions, GPz-Normal shown in red (solid line), DJINN in green (dotted).}
\label{fig_synthetic}
\end{figure}

Thus far we have explored the applicability of GPz to building ICF surrogates in the context of the Simplest Design, as a possible pathway to extremely robust implosions. Other ICF design space data sets in general however might have more complex features, in particular i) sharp ignition cliffs, ii) higher dimensionality and iii) larger numbers of simulations. Future work will investigate more fully the performance of GPz in more complex design spaces. Here however we briefly consider a simple analytic model, to see if GPz is likely to have the capacity to scale well (c.f. the simple analytic model considered in Section 2 of \cite{Humbird2018}). We consider a 10-dimensional hypercube with sides going from 0 to 1, and sample 100,000 points randomly in this volume (each dimension sampled uniformly independently). We calculate a mock-yield of $Y=r^{5} (1+100000 \times (1+\mathrm{erf}(10 \times (r-2))))$, where $r$ is the radius from the origin in this 10D space (the error function giving an ignition-like cliff). We then train GPz (with $m=100$ basis functions; same settings as Table \ref{table-settings} used except heteroscedastic=false) and DJINN on 90\% of this data (90,000 points) and test on the other 10\% (10,000 points); results shown in figure \ref{fig_synthetic}. Training time was 415s for GPz and 450s for DJINN. Predicting time was 0.1s for GPz and 7s for DJINN.  Both capture the design space well, with each doing slightly better or worse at different aspects of the prediction e.g. GPz has a few outliers which DJINN doesn't, but it is less biased for high and low $r$. Which statistical properties are most desirable are in general likely to depend on the specific science goal at hand. In summary it appears that GPz should be able to perform well, and run in reasonable amounts of time, for higher dimensional spaces, with larger quantities of data, when there are steep cliffs in the design space.

\section{Conclusions}

Our results show that Gaussian processes have the potential to be useful ICF surrogates, and that in particular GPz is shown to be effective for the task as an easy-to-use algorithm that can cope with huge amounts of data in a high number of dimensions, with realistic uncertainties. In particular GPz may be useful for modelling shot-to-shot variation alongside uncertainty from lack of data in integrated experiments.

Future work will seek to combine experiment and simulation, either through transfer learning as per \cite{Humbird2018}, or possibly through scaling the surrogate as per \cite{Gopalaswamy2019}, taking care to understand biases induced by differences between the target distribution and the training set.

\textbf{Key findings}:
\begin{itemize}
  \item GPz can make highly effective surrogate models for predicting the outcomes ICF experiments
  \item Cost-sensitive learning can help the improve the statistical properties of predictions in the most important parts of parameter space
  \item It is possible to distinguish between uncertainty from shot-to-shot variation and uncertainty from lack of data
  \item The Simplest Design should be able to produce yields of order $10^{15}$ neutrons extremely robustly on NIF
\end{itemize}

\section*{Acknowledgment}

P.W.H. and S.J.R. acknowledge funding from the Engineering and Physical Sciences Research Council. I.A.A acknowledges the support of King Abdulaziz City for Science and Technology. Many thanks to Warren Garbett at the Atomic Weapons Establishment, Aldermaston, Kris McGlinchey and Jeremy Chittenden at Imperial College, London, Jim Gaffney, Luc Peterson, Kelli Humbird and Brian Spears at Lawrence Livermore National Laboratory, and Chris Bridge at Massachusetts General Hospital for useful discussions.

\newpage 




\bibliographystyle{IEEEtran}
\bibliography{GPz_ICF}
%
%
%

\end{document}